\RequirePackage{lineno} 
\documentclass[aps,prc,twocolumn,showpacs,superscriptaddress,nofootinbib]{revtex4-2} %groupedaddress,runinaddress,unsortedaddress,
\DeclareMathAlphabet{\mathpzc}{OT1}{pzc}{m}{it}
\usepackage{lineno}

\usepackage[T1]{fontenc}

\usepackage{graphicx}
\usepackage{amssymb}
\usepackage{amsmath}
\usepackage{amsthm}
\usepackage[percent]{overpic}
\usepackage{float}
\usepackage{array}
\usepackage{multirow}
\usepackage{epsfig}
\usepackage{verbatim}
\usepackage{xcolor}
\usepackage{wrapfig}
\usepackage{enumitem}

\usepackage[hidelinks,colorlinks=true,allcolors=blue,urlcolor=blue]{hyperref}
\usepackage[most]{tcolorbox}

\usepackage{simplemargins}
\settopmargin{2cm}
\setbottommargin{2.2cm}

\setleftmargin{1.5cm} %default: 1.9mm
\setrightmargin{1.4cm} %default: 1.75mm

\newcommand{\DOI}{https://doi.org/}

\newcommand{\D}{\mathrm{d}}
\newcommand{\x}{\chi}
\newcommand{\p}{\mathcal{P}}

\newcommand{\kut}{{\color{black}\Phi}}

\newcommand{\naslov}{A reconciliation of the Pryce-Ward and Klein-Nishina statistics for semi-classical simulations of annihilation photons correlations}

\begin{document}
%\linenumbers

\title{\naslov}

\author{Petar~\v{Z}ugec} \affiliation{Department of Physics, Faculty of Science, University of Zagreb, Zagreb, Croatia}
\author{Eric~Andreas~Vivoda} \affiliation{Department of Physics, Faculty of Science, University of Zagreb, Zagreb, Croatia}
\author{Mihael~Makek} \affiliation{Department of Physics, Faculty of Science, University of Zagreb, Zagreb, Croatia}
\author{Ivica~Fri\v{s}\v{c}i\'{c}} \affiliation{Department of Physics, Faculty of Science, University of Zagreb, Zagreb, Croatia}

\begin{abstract}

Two photons from the ground state para-positronium annihilation are emitted in a maximally entangled singlet state of orthogonal polarizations. In case of the Compton scattering of both photons the phenomenon of quantum entanglement leads to a measurable increase in the azimuthal correlations of scattered photons, as opposed to a classical description treating the two scattering events as independent. The probability of the scattering of the system of entangled photons is described by the Pryce-Ward cross section dependent on a difference of the azimuthal scattering angles in the fixed coordinate frame, while the independent scattering of single photons is described by the Klein-Nishina cross section dependent on the azimuthal angle relative to each photon's initial polarization. Since the singlet state of orthogonal polarizations is rotationally invariant, it does not carry any physical information on the initial polarizations of the single annihilation photons. In such bipartite state the angular origin for the Klein-Nishina cross section is undefined, making the Pryce-Ward and Klein-Nishina descriptions mutually exclusive. However, semi-classical simulations of the joint Compton scattering of entangled photons -- implementing the Pryce-Ward cross section, but still treating the two photons as separate entities -- can reconcile the Pryce-Ward correlations with the Klein-Nishina statistics for single photons by implementing a modified version of a scattering cross section presented in this work.

\end{abstract}

\maketitle

\section{Introduction}

\begin{figure}[t!]
\centering 
\includegraphics[width=1\linewidth]{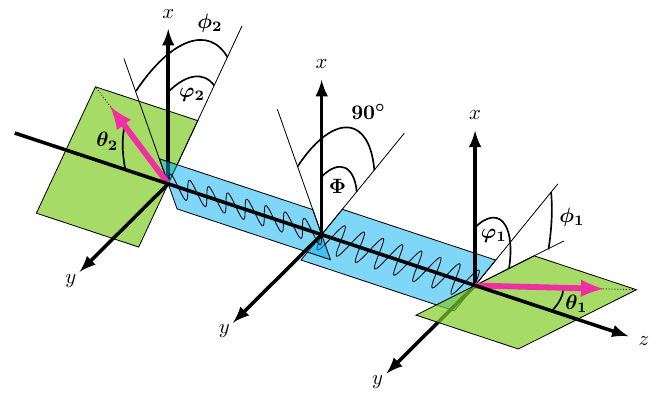}	
\caption{The Compton scattering geometry. Two annihilation photons propagate in the opposite directions along the $z$-axis. Their mutually orthogonal initial polarizations -- indicated by the blue planes -- are here assumed to be well defined. This assumption is the core of a semi-classical treatment. Violet arrows show the scattered photons' momenta, without any indication of their post-scattering polarizations. The green scattering planes provide a spatial indication of the azimuthal scattering angles $\varphi_1,\varphi_2$ relative to the fixed coordinate axis, and of $\phi_1,\phi_2$ relative to the initial photons' polarizations. They are related via one photon's initial polarization angle $\kut$ relative to the fixed axis. For visual purposes the azimuthal angles are shown in a left-handed convention (measured from the $x$-axis towards the negative direction of the $y$-axis).} 
\label{geometry}%
\end{figure}

Two $\gamma$-photons from an annihilation of the ground state para-positronium at rest -- a singlet bound state of electron and positron with anti-parallel spins and the total spin of~0 -- may only be emitted in a Bell state of orthogonal linear polarizations, described by a maximally entangled wavefunction~\cite{aharonov_bohm}:
\begin{equation}
|\Psi\rangle=\frac{1}{\sqrt{2}}\left(|x\rangle_- |y\rangle_+-|y\rangle_- |x\rangle_+\right),
\label{bell_state}
\end{equation}
with $|x\rangle$ and $|y\rangle$ as mutually orthogonal polarization states. Index $\pm$ represents the opposite (back-to-back) directions of photons' propagation, due to the conservation of momentum for a system of a vanishing total momentum (see Fig.~\ref{geometry} for the scattering geometry). One of the remarkable consequences of this entanglement is a non-classical effect on the joint Compton scattering of \text{both} annihilation photons, leading to the Pryce-Ward form of a double-differential cross cross section~\cite{pryce_ward}:
\begin{equation}
\frac{\D^4\sigma_\mathrm{PW}}{\D^2\Omega_1\D^2\Omega_2}=\frac{r_0^4}{16}[F_1F_2-G_1G_2\cos2(\varphi_2-\varphi_1)],
\label{pryce_ward}
\end{equation}
with~$r_0$ as the classical electron radius. The terms $F_i,G_i$ (\mbox{$i=1,2$} for two photons):
\begin{equation}
F_i\equiv \frac{2+(1-\cos\theta_i)^3}{(2-\cos\theta_i)^3} \quad\mathrm{and}\quad G_i\equiv \frac{\sin^2\theta_i}{(2-\cos\theta_i)^2}
\end{equation}
depend on $\theta_i$ as the scattering angles relative to the initial direction of propagation of the annihilation photons. Crucially, $\varphi_i$ are the azimuthal scattering angles \textit{relative to the fixed coordinate frame}, and \mbox{$\D^2\Omega_i=\D(\cos\theta_i)\D\varphi_i$} are the corresponding solid angle elements. The form~(\ref{pryce_ward}) holds only for the 511~keV annihilation photons due to their energy equaling the rest energy $m_ec^2$ of the scattering electrons. Snyder et al.~\cite{snyder} provide a more general form, valid for any photon energy. It is obtained simply by replacing the terms $F_i,G_i$ by their more general forms:
\begin{equation}
F_i= \epsilon_i^3+\epsilon_i-\epsilon_i^2\sin^2\theta_i \quad\mathrm{and}\quad G_i=\epsilon_i^2\sin^2\theta_i,
\end{equation}
valid for any incident photon energy $E_i$, with:
\begin{equation}
\epsilon_i=\frac{E'_i(\theta_i)}{E_i}=\left(1+\frac{E_i}{m_e c^2}(1-\cos\theta_i)\right)^{-1}
\end{equation}
determined by the scattered photon energy $E'_i(\theta_i)$.

Unlike the joint scattering of two entangled photons -- described by the Pryce-Ward cross section -- a scattering of two independent photons is described by the Klein-Nishina cross section~\cite{klein_nishina}:
\begin{equation}
\frac{\D^2\sigma_\mathrm{KN}}{\D^2\omega_i}=\frac{r_0^2}{2}(F_i-G_i\cos2\phi_i),
\label{klein_nishina}
\end{equation}
applying to each photon separately. In particular, the azimuthal angles~$\phi_i$ from the Klein-Nishina formula are defined \textit{relative to each annihilation photon's initial polarization}~\cite{depaola_nima}, hence
the alternative notation for the solid angle element \mbox{$\D^2\omega_i=\D(\cos\theta_i)\D\phi_i$}. Interestingly, only recently has the Compton scattering cross section for single photons been calculated at next-to-leading order in a perturbative QED~\cite{KN_NLO}.

%\vspace*{2cm}

The Pryce-Ward cross section provides a greater azimuthal modulation of scattered photons than can be classically obtained from two independent scatterings based on the Klein-Nishina cross section~\cite{aharonov_bohm,pryce_ward,snyder,caradonna_pra_2024}.  See Refs.~\cite{ivashkin_2023,makek_2024} for a concise overview of the past measurements of this modulation, starting with the first experimental scheme proposed by Wheeler~\cite{wheeler} and the first experimental confirmation by Wu and Shaknow~\cite{wu_shaknow}. A laboratory setup has even been constructed for an undergraduate study of this modulation~\cite{undergraduate}. While the increased modulation (relative to the scattering of separable photons) seems to be beyond doubt -- both theoretically and experimentally -- a question has been recently raised as to whether this modulation serves as the evidence of the initial photons' entanglement~\cite{beatrix}. This was quickly followed by the opposing publication refuting this claim~\cite{caradonna_2019} and soon after by an experimental study supporting it~\cite{exp_study}. Further clarifications were provided in Refs.~\cite{caradonna_pra_2024,lemonade_arxiv}, while subsequent theoretical and experimental studies~\cite{makek_2024,caradonna_2024,decoherence_2024,caradonna_pra_2025,caradonna_pra_2025_x} have moved the discussion well beyond the initial question about the modulation serving as an entanglement witness, toward a fuller understanding of how the entanglement evolves under Compton interactions.

A strong correlation due to the initially entangled polarizations has been recognized as a feature of practical utility, opening a possibility of reducing the noise in the Positron Emission Tomography (PET)~\cite{pet_mcnamara,pet_toghyani,pet_kim,pet_eslami}. A goal of constructing a PET system profiting from such noise reduction is driving a development of the novel experimental setups for detecting the polarization induced correlations~\cite{pet_kim,watts_nature,makek_2020,ana_marija_2021,makek_2022,ana_marija_2024}. The experimental results are already shedding some light on the theoretical issues such as the so called decoherence puzzle, related to a loss of the initial photons' entanglement during the consequent Compton scattering~\cite{caradonna_pra_2024,ivashkin_2023,makek_2024,caradonna_2024,decoherence_2024,caradonna_pra_2025,caradonna_pra_2025_x,watts_nature,sharma,entropy,tkachev}. Another issue starting to be resolved by the experimental studies is the question of the non-maximal entanglement during the ortho-positronium pick-off annihilation in the material, when the positron annihilates with the neighbouring atom-bound electron instead of the electron bound to the positronium~\cite{nonmax_2025}. We mention in passing a related field of active study -- also with the potential for application in PET -- of the entanglement effects in a 3-photon annihilation from ortho-positronium, a triplet bound state of the electron and positron with parallel spins and the total spin of~1~\cite{beatrix,ortho_1,ortho_2,ortho_3,moskal_ortho_ps}.

Due to the complexity of most of the experimental setups (finite acceptances, possible non-uniform detection efficiencies, noise, etc.), a simple analytical formula such as Eq.~(\ref{pryce_ward}) is rarely sufficient for a proper interpretation of the experimental results and their comparison with the theoretical predictions. Instead, one needs to simulate the measurements under the assumption that the measured process follows a given theoretical model. The simulated and the measured data are then to be compared and only then can the proper inferences be made.

An important step in simulating the joint Compton scattering of two entangled annihilation photons was made by Watts et al.~\cite{watts_nature}. They proposed a sampling of the Pryce-Ward cross section for the simulation of the post-scattering angular correlations and implemented this procedure in Geant4 particle simulation toolkit~\cite{geant4_1,geant4_2,geant4_3}. Since Geant4 treats the two photons as separate entities, a technical nature of these simulations is fully classical. However, for simplicity we will call the simulations implementing the non-classical Pryce-Ward correlations \textit{semi-classical}. The method by Watts et al. consists in first letting the Geant4 standard routines handle the scattering of both photons independently, based on the Klein-Nishina formalism. If both photons have scattered, one of them is not interfered with, and it keeps the angular scattering parameters $\theta_1,\varphi_1$ assigned to it by Geant4 (by the random sampling of $\theta_1,\phi_1$ from Eq.~(\ref{klein_nishina}), followed by the appropriate rotation \mbox{$\phi_1\to\varphi_1$} into the fixed coordinate frame). The tracking of the second photon is intercepted and its scattering parameters $\theta_2,\varphi_2$ are reassigned according to Eq.~(\ref{pryce_ward}), wherein $\theta_1,\varphi_1$ of the first photon serve as the fixed parameters for the sampling of $\theta_2,\varphi_2$. We will call this method \textit{a direct Pryce-Ward sampling}. An extended approach to simulating any initial bipartite polarization state with Geant4 was recently reported by Ba\l a et al.~\cite{simul_new}.

It should be noted that the knowledge of the initial photon polarizations is incompatible with the maximally entangled state from Eq.~(\ref{bell_state}). Indeed, in this state their specific polarizations are physically undefined, thus making the Pryce-Ward and the Klein-Nishina descriptions simultaneously incompatible. We show in this work that they may be artificially reconciled, with the specific purpose of improving the semi-classical simulations of a joint scattering process. Figure~\ref{geometry} illustrates a relation between the relevant azimuthal angles $\varphi_i$ and $\phi_i$, assuming a semi-classical description where we imagine the initial polarizations of the entangled photons to be well defined.

\section{The issue}
\label{sec_issue}

Though it may be obscured by using a particular basis \mbox{$\{|x\rangle,|y\rangle\}$} of single-photon states polarized along the $x$ and $y$-axis, the maximally entangled singlet state from Eq.~(\ref{bell_state}) is rotationally invariant. Thus, no physical information about the polarization of single photons is available from it. This is readily demonstrated by considering the state $|\kut\rangle$ of an arbitrary linear polarization:
\begin{equation}
|\kut\rangle=\cos\kut |x\rangle +\sin\kut |y\rangle.
\end{equation}
When constructing a singlet bipartite state from the arbitrarily rotated basis \mbox{$\{|\kut\rangle,|\kut+\tfrac{\pi}{2}\rangle\}$} of orthogonal polarizations (see Fig.~\ref{geometry}):
\begin{align}
\begin{split}
%|\kut\rangle_- |\kut+\tfrac{\pi}{2}\rangle_+-|\kut+\tfrac{\pi}{2}\rangle_- |\kut\rangle_+=|x\rangle_- |y\rangle_+-|y\rangle_- |x\rangle_+,
&|\kut\rangle_- |\kut+\tfrac{\pi}{2}\rangle_+-|\kut+\tfrac{\pi}{2}\rangle_- |\kut\rangle_+\\
=&|x\rangle_- |y\rangle_+-|y\rangle_- |x\rangle_+,
\label{rot_invariance}
\end{split}
\end{align}
the same Bell state~$|\Psi\rangle$ is obtained as with the \mbox{$\{|x\rangle,|y\rangle\}$} basis. Since~$|\Psi\rangle$ is entirely invariant of single photons' (assumed) polarizations, it leaves the specific photons' polarizations \textit{entirely undefined}.

Since the initial polarizations are not defined by the Bell state from Eq.~(\ref{bell_state}), no angular origin can be defined for determining the initial-polarization-relative~$\phi_i$ appearing in the Klein-Nishina cross section. Therefore, the Pryce-Ward and the Klein-Nishina description are mutually exclusive. The issue is analogous to the well known double-slit experiment, which we imagine to be performed with electrons.
\begin{itemize}[noitemsep,topsep=2pt,leftmargin=0.05\linewidth]
\item When electrons freely pass through both slits, a non-classical interference pattern appears on the screen behind the slits. This is analogous to the Pryce-Ward description of a joint photon scattering, when no information on specific photon polarizations is available.
\item When the passage of electrons through particular slits is detected (i.e. a particular-slit information is available), a \textit{classical} interference pattern appears on the screen, spoiling a non-classical outcome. This is analogous to the Klein-Nishina description of independent photon scatterings, wherein the detection of (i.e. the very availability of information on) initial photon polarizations destroys the entanglement-generated interference.
\end{itemize}
Thus, the specific information on initial photon polarizations is \textit{in principle} incompatible with the Pryce-Ward correlations. However, in semi-classical simulations such as those by Watts et al.~\cite{watts_nature} and Ba\l a et al.~\cite{simul_new} -- where the annihilation photons are treated as separate entities and the quantum correlations are only imitated by sampling the Pryce-Ward distribution as the quantum outcome known in advance -- the information on the initial photon polarizations \textit{is}, in fact, fully available. This opens a possibility of simultaneously simulating the Pryce-Ward correlations between the two photons, as well as the Klein-Nishina statistics for single photons. This utility is further discussed in Section~\ref{utility}.

However, an entirely non-evident issue appears in implementing a direct Pryce-Ward sampling procedure, which is now a basis for the \textit{entanglement mode} implemented in Geant4 since version~11.0. It consists of the azimuthal part of the second photon's scattering distribution (for the initial-polarization-relative~$\phi_2$) being inconsistent with the first photon's distribution. In place of the Klein-Nishina distribution from Eq.~(\ref{klein_nishina}) for the first photon, a core of the second photon's angular distribution becomes \mbox{$F_2-\lambda G_2\cos\phi_2$}, the azimuthal modulation being suppressed by a factor \mbox{$\lambda\approx0.08457$}.

\begin{figure}[t!]
\centering 
\includegraphics[width=1\linewidth]{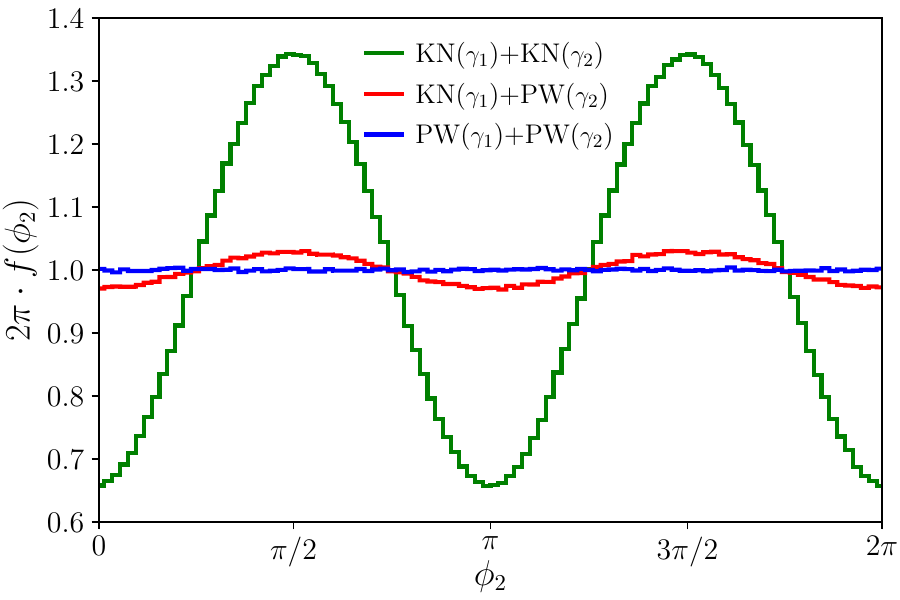}	
\caption{Reduced distributions of the initial-polarization-relative azimuthal scattering angle $\phi_2$ for the second annihilation photon, marginalized over the scattering angle \mbox{$\theta_2\in[0,\pi]$} relative to the photon's initial direction of propagation. Each distribution is obtained by the Monte-Carlo integration, i.e. by a random sampling of the Klein-Nishina (KN) and/or Pryce-Ward (PW) cross section, using $4\cdot10^7$ samples. Labels indicate whether the first~$\gamma_1$ and the second~$\gamma_2$ photon were generated according to the KN or the PW distribution. The analytic form of these distributions is given by Eqs.~(\ref{f_kn_kn}), (\ref{f_kn_pw}), (\ref{f_pw_pw}). The probability distributions $f(\phi_2)$ are scaled by a factor $2\pi$.} 
\label{kn_vs_pw}%
\end{figure}

The issue is demonstrated in Fig.~\ref{kn_vs_pw}, which was obtained by applying a direct Pryce-Ward sampling procedure. We assume the two annihilation photons $\gamma_1$ and $\gamma_2$ to be emitted in the opposite directions along the $z$-axis of the fixed coordinate frame, with initially perpendicular linear polarizations (see Fig.~\ref{geometry}). First, the two photons' initial polarizations are randomly generated. An angle $\kut$ is uniformly sampled from $[0,2\pi]$ for the first photon's initial polarization relative to the fixed coordinate axis. For the second photon an additional uniform sampling between the two elements (e.g. $\pm1$) is performed in order for its initial polarization to be set ``to the left'' or ``to the right'' from the first photon's polarization, thus assigning either \mbox{$\kut+\pi/2$} or \mbox{$\kut-\pi/2$} to the second photon's initial polarization. The scattering angles $\theta_1,\phi_1$ for $\gamma_1$ are then generated by sampling the Klein-Nishina cross section from Eq.~(\ref{klein_nishina}). This gives us the azimuthal angle $\phi_1$ \textit{relative to the photon's initial polarization}. A simple transformation \mbox{$\varphi_1=(\kut+\phi_1) \;\mathrm{mod}\; 2\pi$} provides the scattering angle \textit{relative to the fixed coordinate axis}. The scattering angles $\theta_2,\varphi_2$ for $\gamma_2$ are generated according to the Pryce-Ward cross section from Eq.~(\ref{pryce_ward}), with $\theta_1,\varphi_1$ of the first photon as fixed parameters. The azimuthal angle $\phi_2$ relative to the second photon's initial polarization is then obtained as \mbox{$\phi_2=(\varphi_2-\kut\mp\pi/2) \;\mathrm{mod}\; 2\pi$}. The red plot from Fig.~\ref{kn_vs_pw} shows thus obtained reduced distribution $f_\mathrm{KN+PW}(\phi_2)$ marginalized over all values of $\theta_2$. The notation `KN' or `PW' symbolically indicates which cross section -- Klein-Nishina or Pryce-Ward -- was used for sampling the scattering parameters for a specific photon. Since the mean height of all (properly normalized) probability distributions from Fig.~\ref{kn_vs_pw} is \mbox{$1/2\pi$}, for visual clarity we show them scaled by a factor of $2\pi$.

For comparison, the green plot shows a distribution $f_\mathrm{KN+KN}(\phi_2)$ when the polarization-relative $\phi_2$ is sampled directly from the Klein-Nishina cross section for the second photon, without utilizing the Pryce-Ward distribution. In both cases the green plot also corresponds to a distribution of $\phi_1$ scattering angles of the first photon. This is how the distribution \textit{should} appear if one were expecting (artificially, due to the possibilities opened by the semi-classical simulations) a successful reconstruction of the single-photon statistics, alongside the two-photon correlations. The form of this reduced distribution is obtained by the most basic probability manipulations:
\begin{equation}
f_\mathrm{KN+KN}(\phi_2)=\frac{1}{2\pi}\left(1-\frac{\mathcal{G}}{\mathcal{F}}\cos2\phi_2\right);
\label{f_kn_kn}
\end{equation}
see Eq.~(\ref{app_f_kn_kn}) from Section~\ref{appendix_kn_kn} of the Supplementary note. Factors \mbox{$\mathcal{F}=\int_{-1}^1 F_i \D(\cos\theta_i)$} and $\mathcal{G}=\int_{-1}^1 G_i \D(\cos\theta_i)$, such that \mbox{$\mathcal{G}/\mathcal{F}\approx0.3434$}, are defined by Eq.~(\ref{ff_gg}) from the Supplementary note. The form of the red distribution from Fig.~\ref{kn_vs_pw} -- obtained by sampling the Klein-Nishina cross section for the first photon and the Pryce-Ward cross section for the second one -- is found by a somewhat more involved calculation carried out in Section~\ref{appendix_kn_pw} of the Supplementary note:
\begin{equation}
f_\mathrm{KN+PW}(\phi_2)=\frac{1}{2\pi}\left(1-\lambda\frac{\mathcal{G}}{\mathcal{F}}\cos2\phi_2\right).
\label{f_kn_pw}
\end{equation}
Its azimuthal modulation is greatly suppressed by a factor \mbox{$\lambda=(2\mathcal{F})^{-1}\int_{-1}^1(G_i^2/F_i)\D(\cos\theta_i)\approx 0.08457$}.

If all four scattering parameters $\theta_1,\varphi_1,\theta_2,\varphi_2$ are sampled at once from the Pryce-Ward cross section -- followed by a transition to the initial-polarization-relative $\phi_1,\phi_2$:
\begin{align}
&\phi_1^{(\kut)}=(\varphi_1-\kut) \;\mathrm{mod}\; 2\pi,
\label{phi_1}\\
&\phi_2^{(\kut,\pm)}=(\varphi_2-\kut\mp\pi/2) \;\mathrm{mod}\; 2\pi,
\label{phi_2}
\end{align}
wherein $\kut$ and a sign in front of $\pi/2$ are uniformly sampled for each photon pair -- then the distribution of $\phi_1,\phi_2$ for both photons is entirely flat, for all scattering angles $\theta_1,\theta_2$. This is shown by the blue plot from Fig.~\ref{kn_vs_pw}, wherein the reduced distribution $f_\mathrm{PW+PW}(\phi_2)$ is also flat:
\begin{equation}
f_\mathrm{PW+PW}(\phi_2)=\frac{1}{2\pi}.
\label{f_pw_pw}
\end{equation}
Section~\ref{appendix_pw_pw} of the Supplementary note gives analytical justification of this result. Evidently, the azimuthal Klein-Nishina modulation is now completely lost.

In summary, if the two photons were separable, both should feature a green distribution from Fig.~\ref{kn_vs_pw}. When they are maximally entangled, these distributions should not physically exist. From a classical simulation package such as Geant4 these distributions may always be extracted, whether or not the user imagines the photons to be entangled or not. Under these circumstances there are only two sensible options for the entangled photons:
\begin{itemize}[noitemsep,topsep=2pt,leftmargin=0.05\linewidth]
\item either the initial-polarization-relative azimuthal distributions for both photons should be flat, since this is the closest one can come to imitating their non-existence;
\item or the distributions for both photons should have the Klein-Nishina form, since this is precisely their form when they \textit{are} well defined. 
\end{itemize}
A direct Pryce-Ward sampling yields neither of these outcomes, leaving -- as an unintended by-product -- the mutually inconsistent azimuthal distributions for the two photons. Since the distribution for the first photon retains the Klein-Nishina form, it is only natural to expect that the second photon's distribution should do the same. We propose here a solution that achieves this consistency.

%\pagebreak

\section{The proposed solution}

We wish to identify an artificially modified form of the Pryce-Ward cross section. It needs to be a function of the initial-polarization-relative azimuthal angles $\phi_1,\phi_2$, such that it simultaneously recovers the two-photon correlations from Eq.~(\ref{pryce_ward}), as well as the single-photon scattering statistics from Eq.~(\ref{klein_nishina}). Its differential form \mbox{$\D^4\sigma/\D^2\omega_1\D^2\omega_2$} is to be expressed over the solid angle elements \mbox{$\D^2\omega_i=\D(\cos\theta_i)\D\phi_i$}, distinguishing it from the original Pryce-Ward form in Eq.~(\ref{pryce_ward}).

To this end we assume an appropriate ansatz comprising the azimuthal terms $\cos2\phi_1$, $\cos2\phi_2$ and \mbox{$\cos2(\phi_2-\phi_1)$}, as motivated by the Pryce-Ward and Klein-Nishina cross sections themselves. We provide the details of this derivation in Section~\ref{supplement_ansatz} of the Supplementary note. In short, we impose the following conditions upon the modified cross section:
\begin{itemize}[noitemsep,topsep=2pt,leftmargin=0.05\linewidth]
\item it must recover the familiar form of the Price-Ward cross section in the fixed coordinate frame;
\item when marginalized over the angular parameters of one photon, the initial-polarization-relative Klein-Nishina form for the second photon must remain;
\item it must have the same form for both photons, i.e. must be symmetric upon the exchange of photon parameters \mbox{$\theta_1,\phi_1\leftrightarrow\theta_2,\phi_2$};
\item it must be non-negative everywhere.
\end{itemize}
The solution that satisfies all these conditions is not unique. In fact, assuming a minimal physically motivated ansatz, two free parameters remain for a set of possible solutions. Adopting the simplest possible combination (both parameters set to~0; see Section~\ref{supplement_ansatz} of the Supplementary note), the following result remains: 
\begin{align}
\begin{split}
\frac{\D^4\sigma}{\D^2\omega_1\D^2\omega_2}=\frac{r_0^4}{16}\big[& F_1F_2+G_1G_2\cos2(\phi_2-\phi_1)\\
&-(F_2G_1\cos2\phi_1+F_1G_2\cos2\phi_2)\big].
\label{master}
\end{split}
\end{align}
Therefore, this is a \textit{recommended} cross section to be sampled in the semi-classical simulations of the joint Compton scattering of two maximally entangled annihilation photons. Sampling the cross section from Eq.~(\ref{master}) not only recovers the Klein-Nishina distribution of the second photon, but it does so regardless of the sampling succession. The same distribution is obtained whether the first-photon parameters $\theta_1,\phi_1$ are sampled first (from its own Klein-Nishina distribution) or if all the angular parameters $\theta_1,\phi_1,\theta_2,\phi_2$ are sampled at once from Eq.~(\ref{master}) (see Section~\ref{supplement_sampling} of the Supplementary note).

The result from Eq.~(\ref{master}) is further supported by a quantum-mechanical calculation carried out in Section~\ref{supplement_quantum} of the Supplementary note. Furthermore, this particular form has an added advantage of a minimum overall deviation from a physically meaningful case of the two separable photons (see a discussion around Eq.~(\ref{delta_bff_bgg}) from the Supplementary note).

The most important aspect of the modified cross section to be understood is its relation to the Pryce-Ward counterpart. For simplicity, let us denote by $\sigma_\varphi(\varphi_1,\varphi_2)$ a differential Pryce-Ward cross section from Eq.~(\ref{pryce_ward}), and by $\sigma_\phi(\phi_1,\phi_2)$ a differential cross section from Eq.~(\ref{master}). It can be said that $\sigma_\phi(\phi_1,\phi_2)$ is the cross section ``in the reference frame of the initial photon polarizations''. The fact that the azimuthal part of $\sigma_\phi(\phi_1,\phi_2)$ is \textit{not} rotationally invariant is justified by the photons' polarizations breaking the rotational symmetry, by defining a preferred direction for the angular origin of $\phi_1,\phi_2$. Provided that the polarizations of the initial photon pairs are isotropically distributed (a distribution of~$\kut$ from Eqs.~(\ref{phi_1}) and~(\ref{phi_2}) being uniform), the Pryce-Ward form $\sigma_\varphi(\varphi_1,\varphi_2)$ corresponds to the polarization-relative $\sigma_\phi(\phi_1,\phi_2)$ \textit{averaged (i.e. smeared) over the initial polarizations} in the fixed coordinate frame:
\begin{equation}
%\sigma_\varphi(\varphi_1,\varphi_2)=\frac{1}{2\pi}\int_0^{2\pi}\sigma_\phi\big(\phi_1^{(\kut)},\phi_2^{(\kut,\pm)}\big)\D\kut.
\sigma_\varphi(\varphi_1,\varphi_2)=\frac{1}{2\pi}\int_0^{2\pi}\D\kut\cdot \sigma_\phi\big(\phi_1^{(\kut)},\phi_2^{(\kut,\pm)}\big).
\label{average}
\end{equation}
This appears as the central relation in both derivations of Eq.~(\ref{master}). In a derivation from ansatz it corresponds to a Pryce-Ward distribution recovery condition from Eq.~(\ref{condition_1}) in the Supplementary note. In a quantum-mechanical calculation it appears in the form of Eq.~(\ref{final_average}) from Section~\ref{supplement_quantum} of the Supplementary note.

\begin{comment}
Let us further parametrize the classically-assumed initial polarization of the first photon in the fixed coordinate frame by~$\kut$, so that the following relations may be established:
\begin{align}
&\phi_1^{(\kut)}=(\varphi_1-\kut) \;\mathrm{mod}\; 2\pi,\\
&\phi_2^{(\kut,\pm)}=(\varphi_2-\kut\mp\pi/2) \;\mathrm{mod}\; 2\pi,
\end{align}
between the relevant pairs $\phi_1,\phi_2$ and $\varphi_1,\varphi_2$ of azimuthal angles (see Fig.~\ref{geometry} and refer to the Supplementary note for further details). 
\end{comment}

\vspace*{-4mm}

\section{Practical utility}
\label{utility}

\vspace*{-2mm}

Let us imagine two separate scenarios: one involving the maximally entangled annihilation photons, the other involving the two separable photons of the same kinematics as the annihilation photons (separable photons presumably emitted by some process other than the annihilation of a singlet para-positronium state). Application of the recommended cross section from Eq.~(\ref{master}) allows for a single round of simulations to simultaneously produce:
\begin{itemize}[noitemsep,topsep=2pt,leftmargin=0.05\linewidth]
\item the correlations between the entangled photons;
\item the single-photon statistics of the separable photons.
\end{itemize}
When the entangled photons' correlations are used, the single-photon statistics must be ignored. Likewise, if thus obtained single-photon statistics are used, the accompanying two-photon correlations must be ignored, as they correspond to the entangled case. 

%Notwithstanding

In conclusion, the application of the recommended cross section not only allows \textit{all physically defined} scattering aspects of entangled photons to be simulated, but also \textit{some} of the aspects of the separable photons to be captured. This is because the semi-classical simulations of the entangled photons' scattering retain the single-photon statistics, which then act as the unphysical degrees of freedom that can accommodate some additional, useful information. A direct Pryce-Ward sampling procedure makes this information inconsistent between the two photons. The procedure here proposed resolves this inconsistency and allows for a more efficient utilization of the semi-classical simulations. On a general note, Eq.~(\ref{master}) demonstrates that some physically incompatible features may be mathematically reconciled, revealing that they are not logically contradictory. Rather, it is just the nature or our Universe that they can not simultaneously coexist.

\vspace*{-5mm}

\section{Conclusions}

\vspace*{-2mm}

We have demonstrated that the semi-classical simulations of the two-photon correlations in a joint Compton scattering of maximally entangled annihilation photons do not preserve the Klein-Nishina statistics for single photons, when based on a direct sampling of the Pryce-Ward cross section. Though the Pryce-Ward and the Klein-Nishina statistics are physically incompatible -- one applying to the entangled photon state, the other to separable photon states -- the semi-classical simulations of the entanglement-induced correlations nevertheless \textit{do} provide a control over the single-photon degrees of freedom, which are otherwise physically precluded by the entanglement. For this reason, one can attempt to reconcile the Pryce-Ward and the Klein-Nishina statistics, exclusively for the purposes of such semi-classical simulations. We have shown that this is indeed possible and can be achieved by adopting a modified version of a joint-scattering cross section. When obtained from appropriate ansatz, the solution is not unique. We provide a recommended form of the cross section to be used -- obtained from the simplest selection of free parameters -- which is further supported by a guided quantum-mechanical calculation. Thus, it is now possible to simultaneously simulate the two-photon correlations due to the entanglement, and the statistics of single photons as separate, independent quanta.

\vspace*{-5mm}

\section*{Acknowledgements}

\vspace*{-2mm}

This work was supported by the Croatian Science Foundation under the project number IP-2022-10-3878. This research was supported by the European Union -- NextGenerationEU through the National Recovery and Resilience Plan 2021-2026 Institutional grant of University of Zagreb Faculty of Science (Poticanje kompetitivnih projekata i vrhunskih znanstvenih publikacija na Fizi\v{c}kom odsjeku -- ProPubFO). We are grateful to P. Caradonna for a useful discussion and insightful comments. 

\vspace*{-4mm}

\def\bibsection{\section*{\refname}}

%==================================================

\clearpage

%=========
\makeatletter 
\def\set@footnotewidth{\onecolumngrid} % da footnote isto budu onecolumn
\makeatother   
%=========

\onecolumngrid

\numberwithin{equation}{section}
\numberwithin{figure}{section}

\renewcommand{\thepage}{S\arabic{page}}
\setcounter{page}{1}

\renewcommand{\thesection}{\Alph{section}}
\setcounter{section}{0}
\renewcommand{\theHsection}{\Alph{section}} %za linkove

\renewcommand{\thesubsection}{\arabic{subsection}}
\renewcommand{\theHsubsection}{\thesubsection} %za linkove

\renewcommand{\theequation}{\thesection\arabic{equation}} %{S\arabic{equation}}
\setcounter{equation}{0}
\renewcommand{\theHequation}{\thesection\arabic{equation}} %za linkove

\renewcommand{\thefigure}{\thesection\arabic{figure}}
\setcounter{figure}{0}
\renewcommand{\theHfigure}{\thesection\arabic{figure}} %za linkove

\setcounter{footnote}{0}

%=================

\begin{center}

\textbf{\LARGE Supplementary note}\\[.5cm]

\textbf{\large \naslov}\\[.5cm]

Petar~\v{Z}ugec$^{1*}$, Eric~Andreas~Vivoda$^1$, Mihael~Makek$^1$, Ivica~Fri\v{s}\v{c}i\'{c}$^1$
\\[.1cm]
{\small
{\itshape
$^1$Department of Physics, Faculty of Science, University of Zagreb, Zagreb, Croatia\\}
$^*$Electronic address: pzugec@phy.hr\\[.5cm]
%(Dated: \today)\\[.5cm]
}

\begin{minipage}{400pt}
\small
This note presents the supplementary material to the main paper. The references to figures and equations not starting with the alphabetical character -- such as~(1) -- refer to those from the main paper, while those starting with the appropriate letter -- e.g.~(A1) -- refer to those from this note.\\
\end{minipage}

\end{center}

\section{Probability distributions}
\label{appendix}

\subsection{Single-photon distributions}
\label{appendix_kn_kn}

For simplicity of notation we first introduce (\mbox{$i=1,2$}):
\begin{equation}
\x_i\equiv \cos\theta_i,
\end{equation}
so that the basic kinematic functions become:
\begin{equation}
F_i\equiv \frac{2+(1-\x_i)^3}{(2-\x_i)^3} \quad\mathrm{and}\quad G_i\equiv \frac{1-\x_i^2}{(2-\x_i)^2}.
\end{equation}
The corresponding integrals are:
\begin{equation}
\mathcal{F}\equiv\int_{-1}^1 F_i \D\x_i=\frac{40-27\ln3}{9} \quad\mathrm{and}\quad \mathcal{G}\equiv\int_{-1}^1 G_i \D \x_i=4(\ln3-1),
\label{ff_gg}
\end{equation}
allowing us to translate the Klein-Nishina cross section from Eq.~(\ref{klein_nishina}) into the single-photon probability distributions:
\begin{equation}
p(\x_i,\phi_i)=\frac{F_i-G_i \cos2\phi_i}{2\pi\mathcal{F}},
\label{app_pi}
\end{equation}
normalized such that \mbox{$\int_{-1}^1\D\x_i\int_0^{2\pi}\D\phi_i \cdot p(\x_i,\phi_i)=1$}. By marginalizing $p(\x_i,\phi_i)$ over $\x_i$:
\begin{equation}
f_\mathrm{KN+KN}(\phi_2)=\int_{-1}^1 p(\x_2,\phi_2)\D\x_2=\frac{1}{2\pi}\left(1-\frac{\mathcal{G}}{\mathcal{F}}\cos2\phi_2\right)
\label{app_f_kn_kn}
\end{equation}
one immediately obtains a distribution~(\ref{f_kn_kn}) from the main text, shown by the green plot in Fig.~\ref{kn_vs_pw}, and obtained therein by the Monte-Carlo integration.

\subsection{Two-photons distribution}
\label{appendix_pw_pw}

In order to demonstrate how the red-plot distribution from Fig.~\ref{kn_vs_pw} comes about, we need to construct a joint scattering probability distribution \mbox{$\p_\phi(\x_1,\phi_1,\x_2,\phi_2)$} according to the Pyrce-Ward cross section from Eq.~(\ref{pryce_ward}), but dependent on the initial-polarization-relative $\phi_1,\phi_2$. This is a distribution that is \textit{implicitly} sampled during a direct Pryce-Ward sampling procedure proposed by Watts et al.\footnote{
D. P. Watts, J. Bordes, J. R. Brown et al., \textit{Photon quantum entanglement in the MeV regime and its application in PET imaging},\linebreak Nat. Commun. 12 (2021) 2646 [\href{https://doi.org/10.1038/s41467-021-22907-5}{{\DOI}10.1038/s41467-021-22907-5}]
} By normalizing the Pryce-Ward cross section to unity (\mbox{$\iint_{4\pi}\iint_{4\pi} \p_\varphi \D^2\Omega_1\D^2\Omega_2=1$}) we first obtain a dependence on the fixed-coordinate-frame angles $\varphi_1,\varphi_2$:
\begin{equation}
\p_\varphi(\x_1,\varphi_1,\x_2,\varphi_2)=\frac{F_1F_2-G_1G_2\cos2(\varphi_2-\varphi_1)}{(2\pi\mathcal{F})^2}.
\label{app_p_varphi}
\end{equation}
For a given azimuthal angle $\kut$ between the first photon's initial polarization and the fixed coordinate axis perpendicular to its initial impulse (see Fig.~\ref{geometry} from the main paper), the relation between its azimuthal scattering angle $\varphi_1$ \textit{relative to the fixed axis} and its azimuthal scattering angle $\phi_1$ \textit{relative to its initial polarization} reads:
\begin{equation}
\varphi_1^{(\kut)}=(\phi_1+\kut) \;\mathrm{mod}\; 2\pi.
\label{varphi1_from_phi1}
\end{equation}
We have immediately introduced a modulo function so that for \mbox{$\phi_1,\kut\in[0,2\pi]$} a resulting $\varphi_1$ is truncated into the same interval. Since the initial photons' polarizations are orthogonal, the second photon's fixed-axis-relative azimuthal scattering angle $\varphi_2$ must be shifted by a $\pi/2$ term either ``to the left'' or ``to the right'' from the first photon, such that \mbox{$\varphi_2=\varphi_1\pm\pi/2$}. (A more correct claim would be that if we are given two orthogonal polarizations, we don't know \textit{which} photon is of which polarization, i.e. if $\pi/2$ angular shift should be assigned to $\varphi_1$ or $\varphi_2$.) The second photon's $\varphi_2$ and its initial-polarization relative $\phi_2$ thus differ by $\kut\pm\pi/2$. A further \mbox{$\varphi_2\in[0,2\pi]$} requirement yields a relation:
\begin{equation}
\varphi_2^{(\kut,\pm)}=(\phi_2+\kut\pm\pi/2) \;\mathrm{mod}\; 2\pi.
\label{varphi2_from_phi2}
\end{equation}
For a given $\kut$ (the first photon's initial polarization) and a given $+/-$ sign (the second photon's initial polarization), a naive transition between the distributions $\p_\varphi$ and $\p_\phi$ reads:
\begin{equation}
\p_\phi^{(\kut,\pm)}(\x_1,\phi_1,\x_2,\phi_2)=\p_\varphi\Big(\x_1,\varphi_1^{(\kut)},\x_2,\varphi_2^{(\kut,\pm)}\Big).
\end{equation}
The only azimuthal function from Eq.~(\ref{app_p_varphi}) is a double-angle cosine, which satisfies \mbox{$\cos2(\alpha \;\mathrm{mod}\; 2\pi)=\cos2\alpha$}. Therefore, within $\p_\varphi$ a modulo function may be safely dropped, leading to:
\begin{equation}
\p_\phi^{(\kut,\pm)}(\x_1,\phi_1,\x_2,\phi_2)=\frac{F_1F_2+G_1G_2\cos2(\phi_2-\phi_1)}{(2\pi\mathcal{F})^2}
\end{equation}
for any $(\kut,\pm)$ pair. A transformation only affects the cosine sign: \mbox{$\cos2(\varphi_2-\varphi_1)=-\cos2(\phi_2-\phi_1)$}, giving the same result in all cases. Since any combination of $(\kut,\pm)$ parameters is equally probable (and sampled as such for purposes of obtaining the red plot from Fig.~\ref{kn_vs_pw}), the overall distribution $\p_\phi$ should formally be obtained by averaging all such possibilities:
\begin{equation}
\p_\phi(\x_1,\phi_1,\x_2,\phi_2)=\frac{1}{4\pi}\int_0^{2\pi}\D\kut \left[\p_\phi^{(\kut,+)}+\p_\phi^{(\kut,-)}\right].
\end{equation}
The front term \mbox{$\frac{1}{4\pi}=\frac{1}{2}\cdot\frac{1}{2\pi}$} appears as a combination of $1/2$ norm due to two distinct $\pm$ possibilities, and of $1/2\pi$ norm due to $\kut$-averaging. Since $\p_\phi^{(\kut,\pm)}$ are independent of $(\kut,\pm)$, it trivially follows:\footnote{
This result leads to the ``undesirable'' $\phi$-dependent form of the joint scattering cross section:
\begin{equation*}
\frac{\D^4\sigma}{\D^2\omega_1\D^2\omega_2}=\frac{r_0^4}{16}[F_1F_2+G_1G_2\cos2(\phi_2-\phi_1)].
\end{equation*}
which does not reproduce the Klein-Nishina statistics for single photons. 
}
\begin{equation}
\p_\phi(\x_1,\phi_1,\x_2,\phi_2)=\frac{F_1F_2+G_1G_2\cos2(\phi_2-\phi_1)}{(2\pi\mathcal{F})^2}.
\label{app_p_phi}
\end{equation}
If we marginalize this distribution over the parameters of one (e.g. first) photon, due to \mbox{$\int_0^{2\pi}\cos2(\phi_2-\phi_1)\D \phi_1=0$} the azimuthal modulation is immediately and completely lost (prime over $p'$ denoting a single-photon distribution different from the ``polarized'' Klein-Nishina distribution):
\begin{equation}
p'(\x_2,\phi_2)=\int_{-1}^1\D\x_1\int_0^{2\pi}\D\phi_1 \cdot \p_\phi(\x_1,\phi_1,\x_2,\phi_2)=\frac{F_2}{2\pi\mathcal{F}}.
\end{equation}
This loss of Klein-Nishina modulation is due to the specific photons' initial polarizations not being well defined in a two-photon singlet state giving rise to the Pryce-Ward distribution. Additionally marginalizing this result over $\x_2$ leaves a completely flat reduced distribution~(\ref{f_pw_pw}) from the main text:
\begin{equation}
f_\mathrm{PW+PW}(\phi_2)=\int_{-1}^1 p'(\x_2,\phi_2)\D\x_2=\frac{1}{2\pi},
\label{app_f_pw_pw}
\end{equation}
shown in Fig.~\ref{kn_vs_pw} by the blue plot, and obtained therein by a simultaneous sampling of all angular parameters $\x_1,\varphi_1,\x_2,\varphi_2$ from a distribution $\p_\varphi$ in Eq.~(\ref{app_p_varphi}), which is equivalent to a simultaneous sampling of $\x_1,\phi_1,\x_2,\phi_2$ from a distribution $\p_\phi$ in Eq.~(\ref{app_p_phi}).

\subsection{Undesirable azimuthal distribution of the second photon}
\label{appendix_kn_pw}

We can now demonstrate how the suppression of azimuthal modulation from the red plot in Fig.~\ref{kn_vs_pw} from the main text comes about. We remind that this plot is obtained by sampling the first photon's scattering parameters $\x_1,\phi_1$ from the Klein-Nishina distribution $p(\x_1,\phi_1)$ in Eq.~(\ref{app_pi}). This is followed by sampling the second photon's scattering parameters $\x_2,\varphi_2$ from the Pryce-Ward distribution $\p_\varphi$ in Eq.~(\ref{app_p_varphi}), parameterized by $\x_1$ and \mbox{$\varphi_1=(\phi_1+\kut) \;\mathrm{mod}\; 2\pi$}. In light of the subsequent transformation \mbox{$\phi_2=(\varphi_2-\kut\mp\pi/2)\;\mathrm{mod}\; 2\pi$}, this is equivalent to a direct sampling of $\x_2,\phi_2$ from a distribution $\p_\phi$ in Eq.~(\ref{app_p_phi}).

A probability \mbox{$\D^2 \mathbb{P}_1(\x_1,\phi_1)$} of first sampling $\x_1,\phi_1$ (i.e. of sampling the values from an infinitesimal intervals $\D\x_1$ and $\D\phi_1$ around $\x_1$ and $\phi_1$) is:
\begin{equation}
\D^2 \mathbb{P}_1(\x_1,\phi_1)=p(\x_1,\phi_1)\D\x_1\D\phi_1.
\label{app_dp1}
\end{equation}
In the same sense, a probability of sampling these $\x_1,\phi_1$, and then $\x_2,\phi_2$ by a subsequent sampling of $\p_\phi$ is:
\begin{equation}
\D^4 \mathbb{P}_{1\:\mathrm{and}\:2}(\x_1,\phi_1,\x_2,\phi_2)=\D^2 \mathbb{P}_1(\x_1,\phi_1)\D^2 \mathbb{P}_{2\:\mathrm{if}\:1}(\x_2,\phi_2),
\label{app_dp12}
\end{equation}
with $\D^2 \mathbb{P}_{2\:\mathrm{if}\:1}(\x_2,\phi_2)$ as a conditional probability of sampling $\x_2,\phi_2$ \textit{if} $\x_1,\phi_1$ were sampled first. This is simply the axiom of conditional probability, in essence equivalent to the Bayes' theorem. We certainly expect:
\begin{equation}
\D^2 \mathbb{P}_{2\:\mathrm{if}\:1}(\x_2,\phi_2)\propto \p_\phi(\x_1,\phi_1,\x_2,\phi_2)\D\x_2\D\phi_2,
\end{equation}
with $\p_\phi$ from Eq.~(\ref{app_p_phi}). However, since the probability distribution for sampling $\x_2,\phi_2$ is now \textit{parameterized} by $\x_1,\phi_1$ (instead of $\x_1,\phi_1$ still acting as random values to be sampled themselves), the appropriate sampling distribution must be normalized over $\x_2,\phi_2$, meaning that:
\begin{equation}
\D^2 \mathbb{P}_{2\:\mathrm{if}\:1}(\x_2,\phi_2)= \frac{\p_\phi(\x_1,\phi_1,\x_2,\phi_2)\D\x_2\D\phi_2}{\int_{-1}^1\D\x'_2\int_0^{2\pi}\D\phi'_2 \cdot \p_\phi(\x_1,\phi_1,\x'_2,\phi'_2)},
\end{equation}
which leads to:
\begin{equation}
\D^2 \mathbb{P}_{2\:\mathrm{if}\:1}(\x_2,\phi_2)=2\pi\mathcal{F} \frac{\p_\phi(\x_1,\phi_1,\x_2,\phi_2)}{F_1}\D\x_2\D\phi_2.
\label{app_dp2}
\end{equation}
Returning Eqs.~(\ref{app_dp1}) and~(\ref{app_dp2}) into Eq.~(\ref{app_dp12}) yields:
\begin{equation}
\frac{\D^4 \mathbb{P}_{1\:\mathrm{and}\:2}}{\D\x_1\D\phi_1\D\x_2\D\phi_2}=2\pi\mathcal{F}\frac{p(\x_1,\phi_1)\p_\phi(\x_1,\phi_1,\x_2,\phi_2)}{F_1}.
\end{equation}
A marginalized distribution for the second photon alone (a prime in $p'$ again denoting a distribution different from the ``polarized'' Klein-Nishina one) can now be obtained as:
\begin{equation}
p'(\x_2,\phi_2)=\int_{-1}^1\D\x_1\int_0^{2\pi}\D\phi_1\cdot \frac{\D \mathbb{P}^4_{1\:\mathrm{and}\:2}}{\D\x_1\D\phi_1\D\x_2\D\phi_2}.
\end{equation}
Integration yields:
\begin{equation}
p'(\x_2,\phi_2)=\frac{F_2-\lambda G_2 \cos2\phi_2}{2\pi\mathcal{F}},
\end{equation}
with:
\begin{equation}
\lambda=\frac{1}{2\mathcal{F}}\int_{-1}^1\frac{G_1^2}{F_1}\D\x_1\approx\frac{5\ln3}{18\pi\mathcal{F}}\approx 0.08457,
\end{equation}
where the analytic integration result is approximate to a great degree of accuracy. Further marginalizing over $\x_2$:
\begin{equation}
f_\mathrm{KN+PW}(\phi_2)=\int_{-1}^1 p'(\x_2,\phi_2)\D\x_2=\frac{1}{2\pi}\left(1-\lambda\frac{\mathcal{G}}{\mathcal{F}}\cos2\phi_2\right)
\label{app_f_kn_pw}
\end{equation}
leaves a reduced distribution~(\ref{f_kn_pw}) from the main text, shown in Fig.~\ref{kn_vs_pw} by the red plot.\\

Following the same procedure for a properly normalized probability distribution \mbox{$\p_\phi(\x_1,\phi_1,\x_2,\phi_2)$} based on a recommended form of the cross section from Eq.~(\ref{master}) [and soon given by Eq.~(\ref{recommended})] leaves in place of \mbox{$\D^2 \mathbb{P}_{2\:\mathrm{if}\:1}(\x_2,\phi_2)$} from Eq.~(\ref{app_dp2}):
\begin{equation}
\D^2 \mathbb{P}_{2\:\mathrm{if}\:1}(\x_2,\phi_2)=\frac{\p_\phi(\x_1,\phi_1,\x_2,\phi_2)}{p(\x_1,\phi_1)}\D\x_2\D\phi_2,
\end{equation}
with $p$ now the Klein-Nishina distribution from Eq.~(\ref{app_pi}), further yielding:
\begin{equation}
\frac{\D^4 \mathbb{P}_{1\:\mathrm{and}\:2}}{\D\x_1\D\phi_1\D\x_2\D\phi_2}=\p_\phi(\x_1,\phi_1,\x_2,\phi_2).
\end{equation}
Marginalizing over $\x_1,\phi_1$:
\begin{equation}
\int_{-1}^1\D\x_1\int_0^{2\pi}\D\phi_1\cdot \p_\phi(\x_1,\phi_1,\x_2,\phi_2)=p(\x_2,\phi_2),
\label{app_corr_marg}
\end{equation}
the Klein-Nishina distribution for the second photon remains. Marginalization from Eq.~(\ref{app_corr_marg}) is exactly the same that would be performed in obtaining a reduced distribution for $\x_2,\phi_2$ if all scattering parameters $\x_1,\phi_1,\x_2,\phi_2$ were sampled from a recommended \mbox{$\p_\phi(\x_1,\phi_1,\x_2,\phi_2)$} at once. This is why both procedures -- sampling all parameters at once or in stages --  are equivalent when using a recommended distribution.

\section{A modified distribution from ansatz}
\label{supplement_ansatz}

By normalizing the initial-polarization-relative dependence of the recommended cross section from Eq.~(\ref{master}) to unity, one obtains the appropriate probability distribution:
\begin{equation}
\p_\phi(\x_1,\phi_1,\x_2,\phi_2)=\frac{F_1 F_2 +G_1 G_2\cos2(\phi_2-\phi_1)-(F_2 G_1\cos2\phi_1+F_1G_2\cos2\phi_2)}{(2\pi\mathcal{F})^2}
\label{recommended}
\end{equation}
such that \mbox{$\iint_{4\pi}\iint_{4\pi} \p_\phi \D^2\omega_1\D^2\omega_2=1$}. Here we show how to obtain this distribution from an appropriate ansatz, demonstrating in the process that the solution is not unique in a sense of satisfying the imposed constraints.

The first, central condition is that the sought candidate(s) for the initial-polarization-relative $\p_\phi$ be able to reproduce the fixed-axes-relative distribution \mbox{$\p_\varphi(\x_1,\varphi_1,\x_2,\varphi_2)$} from Eq.~(\ref{app_p_varphi}) in Section~\ref{appendix_pw_pw}, obtained by normalizing the Pryce-Ward cross section from Eq.~(\ref{pryce_ward}). Assuming that the initial photon polarizations and their $\phi_1,\phi_2$ are well defined -- in line with the nature of semi-classical simulations -- we recognize the form $\p_\varphi$ as the result of a uniform distribution of these initial polarizations in the fixed coordinate frame. The transformations between the initial-polarization-relative $\phi_1,\phi_2$ and the fixed-coordinate-axes relative $\varphi_1,\varphi_2$ -- as the inverses of Eqs.~(\ref{varphi1_from_phi1}) and~(\ref{varphi2_from_phi2}) -- read:
\begin{align}
&\phi_1^{(\kut)}=(\varphi_1-\kut) \;\mathrm{mod}\; 2\pi,
\label{phi_varphi_1}\\
&\phi_2^{(\kut,\pm)}=(\varphi_2-\kut\mp\pi/2) \;\mathrm{mod}\; 2\pi,
\label{phi_varphi_2}
\end{align}
with $\kut$ parameterizing the first photon's initial polarization and any of the terms $\pm\pi/2$ parameterizing the orthogonality of the second photon's initial polarization. The fixed-coordinate-axes relative distribution $\p_\varphi$ should remain upon averaging the initial polarization directions in the fixed coordinate frame, yielding a condition:\footnote{
Strictly speaking, the averaging should also be performed over the two $+/-$ possibilities:
\begin{equation*}
\frac{1}{4\pi}\int_0^{2\pi}\D\kut \cdot \left[\p_\phi \Big(\theta_1,\phi_1^{(\kut)};\theta_2,\phi_2^{(\kut,+)}\Big) +\p_\phi \Big(\theta_1,\phi_1^{(\kut)};\theta_2,\phi_2^{(\kut,-)}\Big) \right] =\p_\varphi(\theta_1,\varphi_1;\theta_2,\varphi_2).
\end{equation*}
However, we anticipate using only the azimuthal functions of the form $\cos2\alpha$, which are insensitive to the involved $+/-$ difference: \mbox{$\cos2\phi_2^{(\kut,\pm)}=-\cos2(\varphi_2-\kut)$}. Thus, the entire $\p_\phi$ should be insensitive to it and the averaging may be performed only over~$\kut$. In that, we actively use the fact that \mbox{$\cos2(\alpha \;\mathrm{mod}\; 2\pi)=\cos2\alpha$}, so that the modulo operation is dropped from the later explicit expressions.
}
\begin{equation}
\frac{1}{2\pi}\int_0^{2\pi}\D\kut \cdot \p_\phi \Big(\theta_1,\phi_1^{(\kut)};\theta_2,\phi_2^{(\kut,\pm)}\Big) =\p_\varphi(\theta_1,\varphi_1;\theta_2,\varphi_2)
\label{condition_1}
\end{equation}
obtained by applying the averaging operator\footnote{
Assuming a uniform angular distribution of initial polarizations means, of course, that \textit{whatever} a distribution of the initial-polarization-relative~$\phi_i$ may be, a distribution of scattering angles~$\varphi_i$ in the fixed coordinate frame will immediately be uniform.
} $\frac{1}{2\pi}\int_0^{2\pi}\D\kut$.

% (see the end-note about the domain of azimuthal integration).

\pagebreak

The second condition is that by marginalizing a sought distribution over all parameters of the second photon, a Klein-Nishina distribution \mbox{$p(\x_1,\phi_1)$} from Eq.~(\ref{app_pi}) for the first photon remains:
\begin{equation}
\int_{-1}^1\D\x_2 \int_0^{2\pi}\D\phi_2 \cdot \p_\phi(\x_1,\phi_1,\x_2,\phi_2)=p(\x_1,\phi_1).
\label{condition_2}
\end{equation}
The third condition is the symmetry upon the exchange \mbox{$\x_1,\phi_1\leftrightarrow \x_2,\phi_2$} of the two photons' parameters:\footnote{
A symmetry condition also ensures that a marginalization over the parameters of the first photon:
\begin{equation*}
\int_{-1}^1\D\x_1 \int_0^{2\pi}\D\phi_1 \cdot \p_\phi(\x_1,\phi_1,\x_2,\phi_2)=p(\x_2,\phi_2)
\end{equation*}
reproduces the Klein-Nishina distribution for the second photon as well.
}
\begin{equation}
\p_\phi(\x_1,\phi_1,\x_2,\phi_2)=\p_\phi(\x_2,\phi_2,\x_1,\phi_1).
\label{condition_3}
\end{equation}
The final condition is that the sought distribution be non-negative:
\begin{equation}
\p_\phi(\x_1,\phi_1,\x_2,\phi_2)\ge0 \quad\mathrm{for}\:
\left\{\begin{array}{l}
\x_1,\x_2\in[-1,1] \\
\phi_1,\phi_2\in[0,2\pi]
\end{array}\right. .
\label{condition_4}
\end{equation}
This condition is more difficult to analyse algebraically, so it is best ensured numerically (\textit{a posteriori}, after the specific distribution candidates have been obtained).
\begin{equation*}
\ast\ast\ast
\end{equation*}

We first consider a minimal ansatz:
\begin{align}
\begin{split}
\p_\phi(\x_1,\phi_1,\x_2,\phi_2) &=  F_1F_2 \big[A_{FF} + B_{FF}\cos2\phi_1+ C_{FF}\cos2\phi_2 + D_{FF}\cos2(\phi_2-\phi_1) \big]\\
&+G_1G_2 \big[A_{GG} + B_{GG}\cos2\phi_1+ C_{GG}\cos2\phi_2+ D_{GG}\cos2(\phi_2-\phi_1) \big]\\
&+F_1G_2 \big[A_{FG} + B_{FG}\cos2\phi_1+ C_{FG}\cos2\phi_2 + D_{FG}\cos2(\phi_2-\phi_1) \big]\\
&+G_1F_2 \big[A_{GF} + B_{GF}\cos2\phi_1+ C_{GF}\cos2\phi_2+ D_{GF}\cos2(\phi_2-\phi_1) \big]
\end{split}
\end{align}
with free coefficients $A_{ij},B_{ij},C_{ij},D_{ij}$ ($i,j\in\{F,G\}$), comprising only those azimuthal-basis functions appearing in the Pryce-Ward and the Klein-Nishina distributions. For visual clarity we can write the same ansatz in the vector-matrix-vector multiplication form:
\begin{equation}
\p_\phi(\x_1,\phi_1,\x_2,\phi_2) = 
\left[ \begin{array}{c} F_1F_2 \\ G_1G_2 \\ F_1G_2 \\ G_1F_2 \end{array} \right]
\left[ \begin{array}{cccc} 
A_{FF} & B_{FF} & C_{FF} & D_{FF} \\
A_{GG} & B_{GG} & C_{GG} & D_{GG} \\
A_{FG} & B_{FG} & C_{FG} & D_{FG} \\
A_{GF} & B_{GF} & C_{GF} & D_{GF} \\
\end{array} \right]
\left[ \begin{array}{c} 1 \\ \cos2\phi_1 \\ \cos2\phi_2 \\ \cos2(\phi_2-\phi_1) \end{array} \right].
\label{ansatz_0}
\end{equation}
The polar-basis functions (dependent on $\x_1,\x_2$) are isolated within the left vector (which is to be understood as transposed, i.e. a row-vector), with the azimuthal-basis functions composing the right column-vector.\\

The conditions from Eqs.~(\ref{condition_1}) and~(\ref{condition_2}) now lead to two equations, here written side by side:
\begin{align}
\begin{split}
& F_1F_2[A_{FF}-D_{FF}\cos2(\phi_2-\phi_1)]\\
+&G_1G_2[A_{GG}-D_{GG}\cos2(\phi_2-\phi_1)]\\
+&F_1G_2[A_{FG}-D_{FG}\cos2(\phi_2-\phi_1)]\\
+&G_1F_2[A_{GF}-D_{GF}\cos2(\phi_2-\phi_1)]\\
=&\frac{F_1F_2-G_1G_2\cos2(\varphi_2-\varphi_1)}{(2\pi\mathcal{F})^2}
\end{split}
\qquad\mathrm{and}\hspace*{-2.3cm}
\begin{split}
&2\pi\mathcal{F}F_1(A_{FF}+B_{FF} \cos2\phi_1)\\
+&2\pi\mathcal{G}G_1(A_{GG}+B_{GG} \cos2\phi_1)\\
+&2\pi\mathcal{G}F_1(A_{FG}+B_{FG} \cos2\phi_1)\\
+&2\pi\mathcal{F}G_1(A_{GF}+B_{GF} \cos2\phi_1)\\
=&\frac{F_1-G_1\cos2\phi_1}{2\pi\mathcal{F}}
\end{split}
\end{align}
with the symmetry condition from Eq.~(\ref{condition_3}) best expressed in a matrix notation:
\begin{align}
\begin{split}
\left[ \begin{array}{c} F_1F_2 \\ G_1G_2 \\ F_1G_2 \\ G_1F_2 \end{array} \right]
\left[ \begin{array}{cccc} 
A_{FF} & B_{FF} & C_{FF} & D_{FF} \\
A_{GG} & B_{GG} & C_{GG} & D_{GG} \\
A_{FG} & B_{FG} & C_{FG} & D_{FG} \\
A_{GF} & B_{GF} & C_{GF} & D_{GF} \\
\end{array} \right]
\left[ \begin{array}{c} 1 \\ \cos2\phi_1 \\ \cos2\phi_2 \\ \cos2(\phi_2-\phi_1) \end{array} \right]
=
\left[ \begin{array}{c} F_1F_2 \\ G_1G_2 \\ G_1F_2 \\ F_1G_2 \end{array} \right]
\left[ \begin{array}{cccc} 
A_{FF} & B_{FF} & C_{FF} & D_{FF} \\
A_{GG} & B_{GG} & C_{GG} & D_{GG} \\
A_{FG} & B_{FG} & C_{FG} & D_{FG} \\
A_{GF} & B_{GF} & C_{GF} & D_{GF} \\
\end{array} \right]
\left[ \begin{array}{c} 1 \\ \cos2\phi_2 \\ \cos2\phi_1 \\ \cos2(\phi_2-\phi_1) \end{array} \right]
\end{split}.
\end{align}

\pagebreak

\noindent Collecting and comparing the terms multiplying the 16 independent basis functions (for example, the terms next to \mbox{$F_1G_2\cos2\phi_1$} versus those next to \mbox{$F_2G_1\cos2\phi_2$}) yields a series of linear equations for the coefficients $A_{ij},B_{ij},C_{ij},D_{ij}$. 

\begin{center}
\textit{The solution exists, but is not unique.}
\end{center}

\noindent In fact, two free coefficients remain, for which we select $B_{FF}$ and $B_{GG}$. The general solution (under the assumption of minimal ansatz) reads:
\begin{align}
\begin{split}
\p_\phi(\x_1,\phi_1,\x_2,\phi_2)&= \frac{F_1 F_2 +G_1 G_2\cos2(\phi_2-\phi_1)-[F_2 G_1\cos2\phi_1+F_1G_2\cos2\phi_2]}{(2\pi\mathcal{F})^2}\\
&+\frac{(\mathcal{G}F_2-\mathcal{F}G_2)(B_{FF}\mathcal{F}F_1-B_{GG}\mathcal{G}G_1)\cos2\phi_1 + (\mathcal{G}F_1-\mathcal{F}G_1)(B_{FF}\mathcal{F}F_2-B_{GG}\mathcal{G}G_2)\cos2\phi_2}{\mathcal{F}\mathcal{G}}.
\end{split}
\label{P_BFF_BGG}
\end{align}
One still needs to ensure that a solution (i.e. a combination of $B_{FF},B_{GG}$) exists such that the distribution is nowhere negative, as per constraint from Eq.~(\ref{condition_4}). Indeed, a continuum of such solutions exists, though a parameter subspace for $B_{FF},B_{GG}$ which yield the entirely non-negative solutions is quite narrow. Selecting \mbox{$B_{FF},B_{GG}=0$} provides the simplest possible solution from Eq.~(\ref{recommended}), corresponding to the recommended cross section from Eq.~(\ref{master}).
\begin{equation*}
\ast\ast\ast
\end{equation*}

Any selection of free coefficients $B_{FF},B_{GG}$ -- consistent with the distribution non-negativity condition~(\ref{condition_4}) -- reproduces the single-photon Klein-Nishina statistics and the Pryce-Ward correlations equally well. As such, a selection \mbox{$B_{FF},B_{GG}=0$} is the most convenient one for practical implementation, as it yields the simplest analytical form by eliminating the entire second fractional term from Eq.~(\ref{P_BFF_BGG}). This selection if also favoured by a guided quantum-mechanical calculation from Section~\ref{supplement_quantum}. Furthermore, it can be said that \mbox{$B_{FF},B_{GG}=0$} is ``minimally biased'', in a sense of a minimal deviation from a physically meaningful case of the two separable photons. Let us denote the joint distribution of the two separable photons as $\p_\phi^{(\mathrm{KN})}$, as it is just the product of the two photons' Klein-Nishina distributions from Eq.~(\ref{app_pi}):
\begin{equation}
\p_\phi^{(\mathrm{KN})}(\x_1,\phi_1,\x_2,\phi_2)=\frac{(F_1-G_1 \cos2\phi_1)(F_2-G_2 \cos2\phi_1)}{(2\pi\mathcal{F})^2}.
\end{equation}
For purposes of this discussion let us denote the general solution from Eq.~(\ref{P_BFF_BGG}) as $\p_\phi^{(B_{FF},B_{GG})}$, since we are now interested in its dependence on $B_{FF},B_{GG}$. We can now observe a dependence $\Delta(B_{FF},B_{GG})$ of the $L^2$-distance between the two distributions (akin to the ``root mean square'' of the deviations between them):
\begin{equation}
\Delta(B_{FF},B_{GG})=\int_{-1}^1\D\x_1 \int_{-1}^1\D\x_2 \int_0^{2\pi}\D\phi_1 \int_0^{2\pi}\D\phi_2 \cdot \left[\p_\phi^{(B_{FF},B_{GG})} - \p_\phi^{(\mathrm{KN})}\right]^2.
\label{delta_bff_bgg}
\end{equation}
The result is of the form:
\begin{equation}
\Delta(B_{FF},B_{GG})=a B_{FF}^2+ b B_{GG}^2+ c B_{FF}B_{GG} +d,
\end{equation}
with long and tedious expressions for pure numbers $a,b,c,d$, numerically amounting to \mbox{$a\approx15.9$}, \mbox{$b\approx0.2$}, \mbox{$c\approx-2.7$}, \mbox{$d\approx 3.6\cdot10^{-5}$}. With that, it can easily be shown that the $L^2$-distance has a minimum precisely at \mbox{$B_{FF},B_{GG}=0$}, meaning that this selection of parameters provides a minimal overall deviation of the recommended joint distribution from the physical case of two separable photons.
\begin{equation*}
\ast\ast\ast
\end{equation*}
%(nema smisla uspoređivati odstupanje od Pryce-Warda jer je on u krivim varijablama [kutovima iz fixnog sustava])

%admissible
%boasts 

%\clearpage

\begin{figure}[t!]
\centering 
\includegraphics[width=0.4\linewidth]{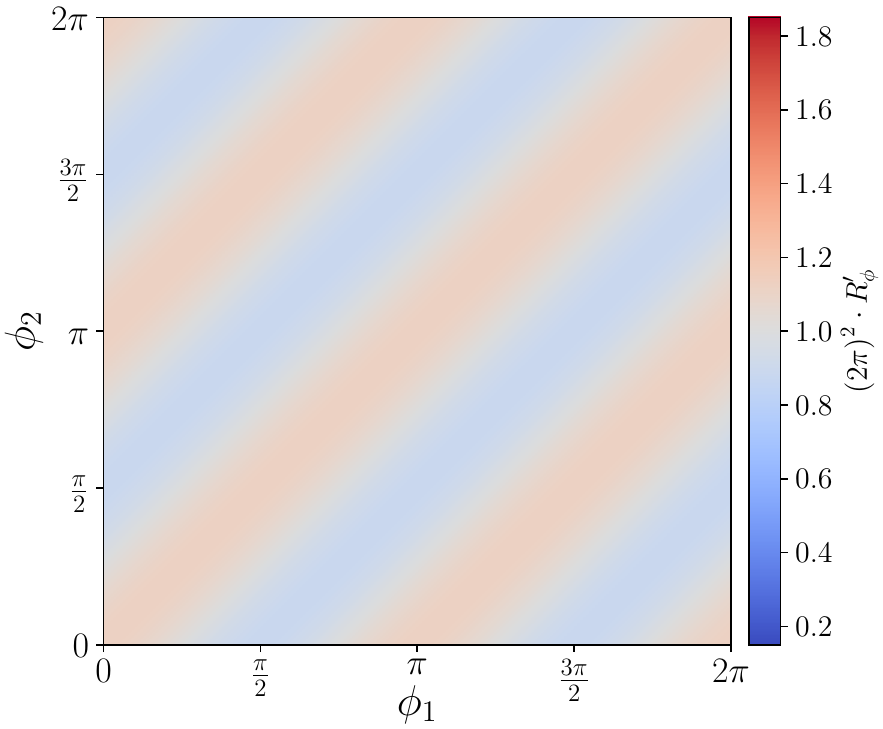}\hspace*{5mm}\includegraphics[width=0.4\linewidth]{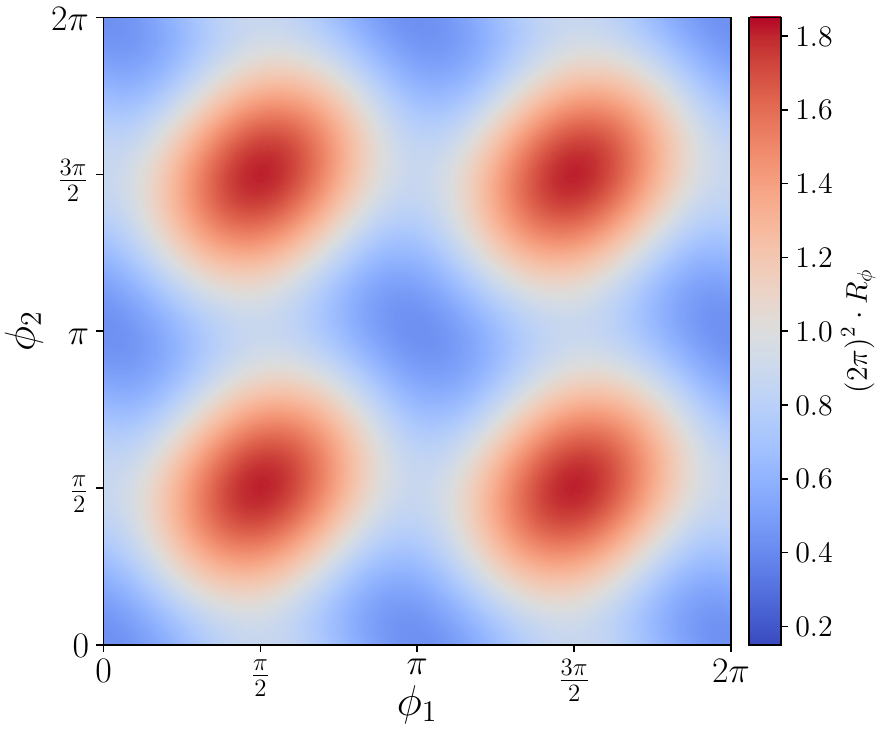}
\caption{Reduced (marginalized over $\x_1$ and $\x_2$) distributions from Eqs.~(\ref{reduced1}) and~(\ref{reduced2}). See the main text for details.} 
\label{comparison}%
\end{figure}

Let us denote by $\p'_\phi$ the ``undesirable'' form of the initial-polarization-relative ($\phi$-dependent) scattering distribution from Eq.~(\ref{app_p_phi}). Figure~\ref{comparison} compares this distribution marginalized over $\x_1$ and $\x_2$, i.e. a reduced distribution $R'_\phi$:
\begin{equation}
R'_\phi(\phi_1,\phi_2)=\int_{-1}^1\D\x_1 \int_{-1}^1\D\x_2 \cdot \p'_\phi(\x_1,\phi_1,\x_2,\phi_2)=
\frac{1}{(2\pi)^2}\left[1+\left(\tfrac{\mathcal{G}}{\mathcal{F}}\right)^2 \cos2(\phi_2-\phi_1) \right]
\label{reduced1}
\end{equation}
against the marginalized distribution $R_\phi$ of the recommended scattering distribution $\p_\phi$ from Eq.~(\ref{recommended}):
\begin{equation}
R_\phi(\phi_1,\phi_2)=\int_{-1}^1\D\x_1 \int_{-1}^1\D\x_2 \cdot \p_\phi(\x_1,\phi_1,\x_2,\phi_2)=
\frac{1}{(2\pi)^2} \left[1+\left(\tfrac{\mathcal{G}}{\mathcal{F}}\right)^2 \cos2(\phi_2-\phi_1) - \tfrac{\mathcal{G}}{\mathcal{F}} (\cos2\phi_1+\cos2\phi_2) \right].
\label{reduced2}
\end{equation}
For clarity the color scales show these two distributions scaled by $(2\pi)^2$. Both color scales span the same range so as to visually emphasize how pronounced is the effect of the additional $\cos2\phi_1$, $\cos2\phi_2$  terms relative to the $\cos2(\phi_2-\phi_1)$ term. In that, the displayed $(2\pi)^2 R'_\phi$ distribution from the left plot peaks at \mbox{$1+(\mathcal{G}/\mathcal{F})^2\approx1.118$}, while the $(2\pi)^2 R_\phi$ distribution from the right plot peaks at \mbox{$(1+\mathcal{G}/\mathcal{F})^2\approx1.805$}.

%\pagebreak

\noindent
\begin{equation*}
\ast\ast\ast
\end{equation*}
We see that the solution satisfying the constraints from Eqs.~(\ref{condition_1})--(\ref{condition_4}) is far from unique, even if the minimal ansatz is assumed. The reasoning from Klein-Nishina formula suggests that there is no physical justification for any other set of azimuthal basis functions. However -- purely mathematically -- the set of possible solutions satisfying the imposed constrains is even larger, so that even more general ansatz can be assumed. For example, consider an extended ansatz:
\begin{equation}
\p_\phi(\x_1,\phi_1,\x_2,\phi_2) = 
\left[ \begin{array}{c} F_1F_2 \\ G_1G_2 \\ F_1G_2 \\ G_1F_2 \end{array} \right]
\left[ \begin{array}{ccccccc} 
A_{FF} & B_{FF} & C_{FF} & D_{FF} & X_{FF} & Y_{FF} & Z_{FF} \\
A_{GG} & B_{GG} & C_{GG} & D_{GG} & X_{GG} & Y_{GG} & Z_{GG} \\
A_{FG} & B_{FG} & C_{FG} & D_{FG} & X_{FG} & Y_{FG} & Z_{FG} \\
A_{GF} & B_{GF} & C_{GF} & D_{GF} & X_{GF} & Y_{GF} & Z_{GF} \\
\end{array} \right]
\left[ \begin{array}{c} 1 \\ \cos2\phi_1 \\ \cos2\phi_2 \\ \cos2(\phi_2-\phi_1) \\ \cos2\phi_1\cos2\phi_2 \\ \cos2\phi_1\cos2(\phi_2-\phi_1) \\ \cos2\phi_2\cos2(\phi_2-\phi_1) \end{array} \right].
\label{ansatz_x}
\end{equation}
This time \textit{nine free parameters} remain, parameterizing the possible solutions (a convenient selection consisting of $B_{FF},B_{GG},B_{FG},B_{GF},X_{FF},X_{GG},X_{FG},Y_{FF},Y_{GG}$). We do not pursue this further.

\section{A modified cross section from a quantum-mechanical calculation}
\label{supplement_quantum}

Starting from the photon(s) density matrix \mbox{$\boldsymbol{\rho}$}, a differential scattering cross section may be calculated from a trace \mbox{$\mathrm{Tr}[\boldsymbol{\rho}\boldsymbol{\mathcal{S}}]$}, with $\boldsymbol{\mathcal{S}}$ as an appropriate scattering matrix. For pure (either single-photon or multi-photon) states $|\Psi\rangle$, such that \mbox{$\boldsymbol{\rho}=|\Psi\rangle\langle\Psi|$}, this is equivalent to:
\begin{equation}
\mathrm{Tr}[\boldsymbol{\rho}\boldsymbol{\mathcal{S}}]=\langle\Psi|\boldsymbol{\mathcal{S}}|\Psi\rangle.
\end{equation}
We will use the right hand side expression. For 511~keV annihilation photons, a single-photon scattering matrix $\boldsymbol{\mathcal{S}}_i$ (for the $i$-th photon) takes the form:
\begin{equation}
\boldsymbol{\mathcal{S}}_i=\frac{r_0^2}{2}\mathbf{S}_i =\frac{r_0^2}{2}\big( F_i \mathbb{I} -G_i\cos2\varphi_i \boldsymbol{\sigma}_z-G_i\sin2\varphi_i\boldsymbol{\sigma}_x \big),
\end{equation}
with $r_0$ as the classical electron radius, $\mathbb{I}$ as a $2\times2$ identity matrix and $\boldsymbol{\sigma}_i$ as conventional Pauli matrices. For convenience we have introduced a dimensionless scattering matrix $\mathbf{S}_i$ to be used in the following demonstrations.

\subsection{Single-photon case}
\label{quantum_one_photon}

We first demonstrate that for an arbitrary single-photon polarization state ($\kut$ being a polarization direction in the fixed coordinate frame):
\begin{equation}
|\kut\rangle=\cos\kut |x\rangle +\sin\kut |y\rangle,
\label{phi_rastav}
\end{equation}
the prescribed procedure:
\begin{equation}
\frac{\D^2\sigma_\mathrm{KN}}{\D^2\Omega_i}=\frac{r_0^2}{2}\langle\kut|\mathbf{S}_i|\kut\rangle
\label{sig_kn}
\end{equation}
reproduces a familiar form of the Klein-Nishina cross section. At this point a differential cross section is defined over the solid angle \mbox{$\D^2\Omega_i=\D(\cos\theta_i)\D\varphi_i$} in the fixed-coordinate-frame-relative~$\varphi_i$. For \mbox{$\alpha,\beta\in\{x,y\}$} let us define:
\begin{equation}
S^{\alpha\beta}_i\equiv \langle \alpha|\mathbf{S}_i|\beta\rangle.
\end{equation}
Plugging~$|\kut\rangle$ into Eq.~(\ref{sig_kn}), the following terms appear:
\begin{align}
&S^{xx}_i\equiv \langle x|\mathbf{S}_i|x\rangle = F_i - G_i\cos2\varphi_i,
\label{Sxx}\\
&S^{yy}_i\equiv \langle y|\mathbf{S}_i|y\rangle = F_i + G_i\cos2\varphi_i,
\label{Syy}\\
&S^{xy}_i\equiv \langle x|\mathbf{S}_i|y\rangle = - G_i\sin2\varphi_i,
\label{Sxy}\\
&S^{yx}_i\equiv \langle y|\mathbf{S}_i|x\rangle = - G_i\sin2\varphi_i,
\label{Syx}
\end{align}
so that:
\begin{align}
\begin{split}
\langle\kut|\mathbf{S}_i|\kut\rangle&=S^{xx}_i\cos^2\kut + S^{yy}_i\sin^2\kut +(S^{xy}_i+S^{yx}_i)\sin\kut\cos\kut \\
&=F_i-G_i \cos2(\varphi_i-\kut)=F_i-G_i \cos2\phi_i,
\end{split}
\end{align}
The final transition to the initial-polarization-relative~$\phi_i$ follows from \mbox{$\phi_i=(\varphi_i-\kut) \;\mathrm{mod}\; 2\pi$}. It only remains to make a transition from a differential solid angle \mbox{$\D^2\Omega_i=\D(\cos\theta_i)\D\varphi_i$} to a \mbox{$\D^2\omega_i=\D(\cos\theta_i)\D\phi_i$} in a newly introduced~$\phi_i$. Due to a trivial relationship between $\varphi_i$ and $\phi_i$ (and the fact that they appear as angular \textit{measures}, not requiring a Dirac $\delta$-function during a differentiation), the corresponding differentials transform as \mbox{$\D\varphi_i=\D\phi_i$}, straightforwardly yielding \mbox{$\D^2\Omega_i=\D^2\omega_i$}. Thus, Eq.~(\ref{sig_kn}) finally gives:
\begin{equation}
\frac{\D^2\sigma_\mathrm{KN}}{\D^2\omega_i}=\frac{r_0^2}{2}(F_i-G_i \cos2\phi_i),
\label{kn_omega}
\end{equation}
which is a well established Klein-Nishina result.

\subsection{Two-photons case}
\label{quantum_two_photons}

A calculation of the Pryce-Ward cross section proceeds from a maximally entangled bipartite state:
\begin{equation}
|\Psi\rangle=\frac{1}{\sqrt{2}}\big(|x\rangle_1 |y\rangle_2-|y\rangle_1 |x\rangle_2\big).
\label{psi_state}
\end{equation}
Indices 1,2 designate the two photons' contributions to the wavefunction. The cross section can now be calculated from a scattering matrix $\boldsymbol{\mathcal{S}}$ such that:
\begin{equation}
\boldsymbol{\mathcal{S}}=\frac{r_0^4}{16}\mathbf{S}_1 \otimes \mathbf{S}_2 \quad\Rightarrow\quad \frac{\D^4\sigma_\mathrm{PW}}{\D^2\Omega_1\D^2\Omega_2} = \frac{r_0^4}{16}\langle \Psi|\mathbf{S}_1\otimes \mathbf{S}_2|\Psi\rangle.
\end{equation}
%so that:
%\begin{equation}
%\frac{\D^4\sigma_\mathrm{PW}}{\D^2\Omega_1\D^2\Omega_2} = \frac{r_0^4}{16}\langle \Psi|\mathbf{S}_1\otimes \mathbf{S}_2|\Psi\rangle.
%\end{equation}
Proceeding with the calculation directly from Eq.~(\ref{psi_state}) yields:
\begin{align}
\begin{split}
\langle \Psi|\mathbf{S}_1\otimes \mathbf{S}_2|\Psi\rangle&=\frac{1}{2}(S^{xx}_1 S_2^{yy} + S^{yy}_1 S_2^{xx} -S^{xy}_1 S_2^{yx} - S^{yx}_1 S^{xy}_2)\\
&=F_1F_2 -G_1G_2\cos2(\varphi_2-\varphi_1)
\end{split}
\end{align}
which is the familiar Pryce-Ward form that in no way provides an angular dependence on the initial photons' polarizations, since they are physically undefined in a maximally entangled state [see Eq.~(\ref{rot_invariance}) from the main text].
\begin{equation*}
\ast\ast\ast
\end{equation*}

\pagebreak

Our goal is to gain a cross section dependent on the initial-polarization-relative $\phi_i$, even if by artificial means. To this end we decompose an entangled state $|\Psi\rangle$ into a continuous superposition of bipartite separable states \textit{$|\kut\rangle_1 |\kut\pm\tfrac{\pi}{2}\rangle_2 $}, describing the two photons of well defined orthogonal polarizations:
\begin{equation}
|\Psi\rangle =\int_0^{2\pi} \D\kut \cdot f_{\kut}^{(\pm)} |\kut\rangle_1 |\kut~\pm\tfrac{\pi}{2}\rangle_2.
\label{decomp}
\end{equation}
We simultaneously handle two cases: the second photon's polarization either being ``to the right'' (\mbox{$\kut+\tfrac{\pi}{2}$}) or ``to the left'' (\mbox{$\kut-\tfrac{\pi}{2}$}) from the first photon's. The sought decomposition coefficients $f_{\kut}^{+}$ or $f_{\kut}^{-}$ must be of constant modulus:
\begin{equation}
\big|f_{\kut}^{(\pm)}\big|=\mathrm{const.}
\end{equation}
Otherwise, if any of the specific-polarization states provided a disproportionate contribution through a nonconstant $|f_{\kut}^{(\pm)}\big|$ dependence, the overall $|\Psi\rangle$ would not be rotationally invariant due to the appearance of a preferred direction.

Expressing the separable states via Eq.~(\ref{phi_rastav}):
\begin{align}
\begin{split}
|\kut\rangle_1 |\kut\pm\tfrac{\pi}{2}\rangle_2 &=\pm \big(\cos\kut |x\rangle_1 +\sin\kut |y\rangle_1 \big) \big(-\sin\kut |x\rangle_2 +\cos\kut |y\rangle_2 \big)\\
&=\pm \big( -\sin\kut\cos\kut |x\rangle_1 |x\rangle_2 +\sin\kut\cos\kut |y\rangle_1 |y\rangle_2 +\cos^2\kut |x\rangle_1 |y\rangle_2 -\sin^2\kut |y\rangle_1 |x\rangle_2 \big)
\label{separable}
\end{split}
\end{align}
and comparing their contributions to the target state from Eq.~(\ref{decomp}):
\begin{align}
\begin{split}
\frac{1}{\sqrt{2}}\left(|x\rangle_1 |y\rangle_2-|y\rangle_1 |x\rangle_2\right)=\pm \bigg[ & -\left(\int_0^{2\pi} \D\kut \cdot f_{\kut}^{(\pm)}\sin\kut\cos\kut\right) |x\rangle_1 |x\rangle_2 + \left(\int_0^{2\pi} \D\kut \cdot f_{\kut}^{(\pm)}\sin\kut\cos\kut\right) |y\rangle_1 |y\rangle_2 \\
& +\left(\int_0^{2\pi} \D\kut \cdot f_{\kut}^{(\pm)}\cos^2\kut\right) |x\rangle_1 |y\rangle_2 -\left(\int_0^{2\pi} \D\kut \cdot f_{\kut}^{(\pm)}\sin^2\kut\right) |y\rangle_1 |x\rangle_2\bigg]
\end{split}
\end{align}
yields the following conditions upon the sought $f_{\kut}^{(\pm)}$:
\begin{align}
&\int_0^{2\pi} \D\kut \cdot f_{\kut}^{(\pm)}\sin\kut\cos\kut=0,\\
&\int_0^{2\pi} \D\kut \cdot f_{\kut}^{(\pm)}\cos^2\kut=\pm\frac{1}{\sqrt{2}},\\
&\int_0^{2\pi} \D\kut \cdot f_{\kut}^{(\pm)}\sin^2\kut=\pm\frac{1}{\sqrt{2}}.
\end{align}
All these are trivially satisfied by:
\begin{equation}
f_{\kut}^{(\pm)}=\pm\frac{1}{\pi\sqrt{2}}.
\end{equation}
Therefore, we can use any one of the two decompositions:
\begin{equation}
|\Psi\rangle =\pm\frac{1}{\pi\sqrt{2}}\int_0^{2\pi} \D\kut \cdot |\kut\rangle_1 |\kut~\pm\tfrac{\pi}{2}\rangle_2.
\end{equation}
In calculating \mbox{$\langle \Psi|\mathbf{S}_1\otimes \mathbf{S}_2|\Psi\rangle$} we now decompose only one of the state vectors (e.g. the ket $|\Psi\rangle$):
\begin{align}
\begin{split}
\langle\Psi|\mathbf{S}_1\otimes \mathbf{S}_2|\Psi\rangle= \pm\frac{1}{2\pi}
\left[\big(\langle x|_1 \langle y|_2-\langle y|_1 \langle x|_2\big) \mathbf{S}_1\otimes \mathbf{S}_2\int_0^{2\pi} \D\kut \cdot |\kut\rangle_1 |\kut~\pm\tfrac{\pi}{2}\rangle_2\right].
\end{split}
\end{align}
Bringing everything under the integral, expressing $|\kut\rangle_1 |\kut~\pm\tfrac{\pi}{2}\rangle_2$ as in Eq.~(\ref{separable}) and arranging the terms yields:
\begin{align}
\begin{split}
\langle\Psi|\mathbf{S}_1\otimes \mathbf{S}_2|\Psi\rangle= \frac{1}{2\pi}\int_0^{2\pi} \D\kut 
\Big[& -\sin\kut\cos\kut \big(S_1^{xx}S_2^{yx} - S_1^{yx}S_2^{xx}\big)  +\sin\kut\cos\kut \big(S_1^{xy}S_2^{yy} - S_1^{yy}S_2^{xy}\big)\\
&+\cos^2\kut \big(S_1^{xx}S_2^{yy} - S_1^{yx}S_2^{xy}\big) -\sin^2\kut \big(S_1^{xy}S_2^{yx} - S_1^{yy}S_2^{xx}\big) \Big].
\end{split}
\end{align}
From Eqs.~(\ref{Sxx})--(\ref{Syx}) we now have:
\begin{equation}
\langle\Psi|\mathbf{S}_1\otimes \mathbf{S}_2|\Psi\rangle= \frac{1}{2\pi}\int_0^{2\pi} \D\kut 
\big[F_1F_2 -G_1G_2\cos2(\varphi_2-\varphi_1) -F_2G_1\cos2(\varphi_1-\kut) +F_1G_2\cos2(\varphi_2-\kut) \big].
\end{equation}
Transitioning to the initial-polarization-relative~$\phi_i$ via Eqs.~(\ref{varphi1_from_phi1}) and~(\ref{varphi2_from_phi2}) not only resolves the asymmetry in sign in front of $F_1G_2$ and $F_2G_1$, but it also conceals an explicit dependence on~$\kut$:
\begin{equation}
\langle\Psi|\mathbf{S}_1\otimes \mathbf{S}_2|\Psi\rangle= \frac{1}{2\pi}\int_0^{2\pi} \D\kut 
\big[F_1F_2 -G_1G_2\cos2(\phi_2-\phi_1) -F_2G_1\cos2\phi_1 -F_1G_2\cos2\phi_2 \big],
\end{equation}
allowing us to recognize the recommended form of the initial-polarization-relative cross section under the integral. Expressing it as a differential quantity over \mbox{$\D^2\omega_i=\D(\cos\theta_i)\D\phi_i$} -- using the same rationale as with Eq.~(\ref{kn_omega}) -- immediately provides a clear relation between this result and the physically justified Pryce-Ward form:
\begin{equation}
\frac{\D^4\sigma_\mathrm{PW}}{\D^2\Omega_1\D^2\Omega_2}=\frac{1}{2\pi}\int_0^{2\pi} \D\kut \cdot \frac{\D^4\sigma}{\D^2\omega_1\D^2\omega_2},
\label{final_average}
\end{equation}
representing an averaging of $\D^4\sigma/\D^2\omega_1\D^2\omega_2$ over classically-assumed polarization directions of the initial photon pairs.

\section{Sampling the joint distribution}
\label{supplement_sampling}

By sampling the recommended distribution~(\ref{recommended}) one can confirm that it properly reproduces the single-photon Klein-Nishina statistics from Eq.~(\ref{app_pi}) and the Pryce-Ward correlations\footnote{
Taking into account that the fixed-coordinate-frame \mbox{$\varphi_2-\varphi_1$} are related to the sampled initial-polarization-relative $\phi_1,\phi_2$ through \mbox{$\varphi_2-\varphi_1=\phi_2-\phi_1\pm\pi/2$}, and taking care to fill the histograms with the modulo-wrapped sampled values \mbox{$(\phi_2-\phi_1\pm\pi/2)\;\mathrm{mod}\; 2\pi$}.
} from Eq.~(\ref{app_p_varphi}). To this end one can use a simple \textit{rejection sampling} method. We provide here a basic pseudo-code for sampling all four angular parameters \mbox{$\x_1,\x_2,\phi_1,\phi_2$} at once. In that, one can use only the numerator from Eq.~(\ref{recommended}):
\begin{equation}
N(\x_1,\phi_1,\x_2,\phi_2)=F_1 F_2 +G_1 G_2\cos2(\phi_2-\phi_1)-(F_2 G_1\cos2\phi_1+F_1G_2\cos2\phi_2),
\end{equation}
taking care to identify its maximum value within the relevant angular intervals, which equals~$4$. The basic procedure boils down to uniformly and independently sampling $\x_1,\x_2\in[-1,1]$ and $\phi_1,\phi_2\in[0,2\pi\rangle$ -- together with an additional uniform sampling of $r\in[0,4]$ -- and repeating the procedure until $r\le N(\x_1,\phi_1,\x_2,\phi_2)$ is satisfied:\\

\newcommand{\Tab}{{\color{white}.\quad}}

\noindent \Tab \texttt{repeat:}\\
\Tab\Tab \texttt{uniformly sample}\: $\x_1\in[-1,1]$ \\
\Tab\Tab \texttt{uniformly sample}\: $\x_2\in[-1,1]$ \\
\Tab\Tab \texttt{uniformly sample}\: $\phi_1\in[0,2\pi\rangle$ \\
\Tab\Tab \texttt{uniformly sample}\: $\phi_2\in[0,2\pi\rangle$ \\
\Tab\Tab \texttt{uniformly sample}\: $r\in[0,4]$ \\
\Tab \texttt{while}\: $r>N(\x_1,\phi_1,\x_2,\phi_2)$\\

\noindent Thus generated sets (\mbox{$\x_1,\x_2,\phi_1,\phi_2$}) that terminate the loop are distributed according to the the recommended distribution~(\ref{recommended}). Analogous procedure can be used for sampling any other distribution from this work. Since the distribution~(\ref{recommended}) is consistent with the single-photon Klein-Nishina statistics, one can first generate $\x_1,\phi_1$ from Eq.~(\ref{app_pi}) -- for simplicity, using only the numerator \mbox{$n(\x_1,\phi_1)=F_1-G_1 \cos2\phi_1$}, whose maximum equals~2 -- and then generate $\x_2,\phi_2$ from Eq.~(\ref{recommended}), while using $\x_1,\phi_1$ as fixed parameters:\\

\noindent \Tab \texttt{repeat:}\\
\Tab\Tab \texttt{uniformly sample}\: $\x_1\in[-1,1]$ \\
\Tab\Tab \texttt{uniformly sample}\: $\phi_1\in[0,2\pi\rangle$ \\
\Tab\Tab \texttt{uniformly sample}\: $r\in[0,2]$ \\
\Tab \texttt{while}\: $r>n(\x_1,\phi_1)$\\\\
\Tab \texttt{repeat:}\\
\Tab\Tab \texttt{uniformly sample}\: $\x_2\in[-1,1]$ \\
\Tab\Tab \texttt{uniformly sample}\: $\phi_2\in[0,2\pi\rangle$ \\
\Tab\Tab \texttt{uniformly sample}\: $r\in[0,4]$ \\
\Tab \texttt{while}\: $r>N(\x_1,\phi_1,\x_2,\phi_2)$\\

%===================================================

\end{document}